\documentclass[12pt,preprint]{aastex}

\usepackage{graphicx}
\usepackage{amsfonts}
\usepackage{bm}

\newcommand{\thetaE}{\theta_{\rm E}}
\newcommand{\tE}{t_{\rm E}}

\newcommand{\piE}{\pi_{\rm E}}
\newcommand{\umin}{u_{\rm min}}
\begin{document}

\title{Bringing closure to microlensing mass measurement}

%\title{Mass measurement in gravitational microlensing using closure phase}

\author{Neal Dalal\altaffilmark{1}}
\affil{Institute for Advanced Study, Einstein Drive, Princeton NJ 08540}

\author{Benjamin F. Lane}
\affil{Dept. of Geological and Planetary Sciences, MC 150-21,
Caltech, Pasadena CA 91125}

\altaffiltext{1}{Hubble Fellow}

\begin{abstract}
%closure phase is cool
Interferometers offer multiple methods for studying microlensing events
and determining the properties of the lenses.  We investigate the
study of microlensing events with optical interferometers, focusing on
narrow-angle astrometry, visibility, and closure phase.  After
introducing the basics of microlensing and interferometry, we derive
expressions for the signals in each of these three channels.  For
various forecasts of the instrumental performance, we discuss which
method provides the best means of measuring the lens angular Einstein
radius $\thetaE$, a prerequisite for determining the lens mass.  If
the upcoming generation of large-aperture, AO-corrected long baseline
interferometers (e.g.\ VLTI, Keck, OHANA) perform as well as expected,
$\thetaE$ may be determined with signal-to-noise greater than 10 for
all bright events.  We estimate that roughly a dozen events per year
will be sufficiciently bright and have long enough durations to allow
the measurement of the lens mass and distance from the ground.  
We also consider the prospects for a VLTI survey of all bright lensing
events using a Fisher matrix analysis, and find that even without
individual masses, interesting constraints may be placed on the bulge
mass function, although large numbers of events would be required.
\end{abstract}

\section{Introduction}

Gravitational lensing has been used for nearly a decade to study
faint, compact masses in our galaxy.  Although a large number of
microlensing events have been detected, the lens masses and distances
cannot (in most cases) be determined, meaning that only statistical
constraints on the lensing population may be derived from lensing
surveys \citep[e.g.][]{bulge,lmc5}.  The determination of the lens mass
and distance in individual events would be of great utility towards
elucidating the nature of the microlenses.  An example of this
is the claim by \citet{mao} and \citet{bennett} that three
long-duration events are likely massive black holes, with
$M\sim10-30M_\odot$.  Since individual
masses could not be measured for these events, statistical arguments
were employed to support the claim for large masses.  If confirmed,
\citet{agol} have argued that these black holes would represent a new,
significant population of black holes roaming the Galactic disk.

The measurement of mass in microlensing events requires the
determination of two quantities describing the event: (1) the
lens-source relative parallax, $\piE$, and (2) the angular Einstein
radius, $\thetaE$ \citep{gould}.  Measurement of the parallax, $\piE$,
requires the observation of the lensing event from viewpoints
separated by $\gtrsim 1$ AU.  One way to do this is to observe the
event simultaneously from the ground and from a satellite in solar
orbit; the Space Interferometry Mission (SIM) is currently planned to
do precisely this for a number of future microlensing events.  It is
also possible to measure $\piE$ for long-duration events, with event
timescales $\tE\gtrsim$few months, using the Earth's motion around the
Sun to provide a distant vantage point.  Since $\sim15-20\%$ of
lensing events towards the bulge have durations ${\hat t}=2\tE$ longer
than 100 days, with high quality photometry the parallax may be
measured for a significant fraction of events.

The angular Einstein radius is extremely difficult to measure, since
typical values are $\thetaE\sim0.5 (M/0.3M_\odot)^{1/2}$
milliarcsecond (hereafter mas), 
defying resolution by even the largest telescopes.  In very rare cases
\citep[e.g.][]{an}, it is possible to measure $\thetaE$ during a
caustic crossing, however generically $\thetaE$ cannot be resolved by
any single-aperture instrument.  Resolution of sub-milliarcsecond scales
typically requires long baseline interferometers, and future instruments
like SIM should be able to determine $\thetaE$ for many events
\citep{pac98,boden,gouldsalim} using narrow-angle astrometry.   
Unfortunately, SIM will not fly until later than 2009.
Before SIM flies, a number of long baseline, highly sensitive
interferometers will come online, for example the Keck Interferometer
or the Very Large Telescope Interferometer (VLTI).

These ground-based interferometers can study microlensing events in
several different ways, which we will consider in this paper.  
\citet{boden} discussed how Keck and VLTI can
measure $\thetaE$ using narrow-angle astrometry.  
\citet{delplancke} have pointed out that for massive microlenses,
$\thetaE$ becomes comparable to the resolution $\lambda/B\approx 5$
mas for $B=100$m, $\lambda=2.2\mu$m.  This allows the study of
microlensing events via the partial resolution of the lensed images.
One possible method, discussed by \citet{delplancke}, is the measurement
of the decrement in fringe visibility as the microlensed
images become resolved.  In this paper, we also
investigate the use of closure phase to determine
the angular Einstein radius $\thetaE$.  Closure phase is free of
many of the calibration issues afflicting visibility amplitude,
however other concerns do arise.  In the next section, we
provide a review of the basics of microlensing.  The following section
gives an introduction to interferometry and closure phase.  We then
show how closure phase may be used to measure $\thetaE$, and compare
our method to other proposed techniques.  Since both Keck
Interferometer and VLTI have already observed first fringes, this
method promises to be an exciting technique for the determination of
$\thetaE$ and thereby the lens mass, in the next few years.

\section{Review of microlensing}

\begin{figure}[t]
\plottwo{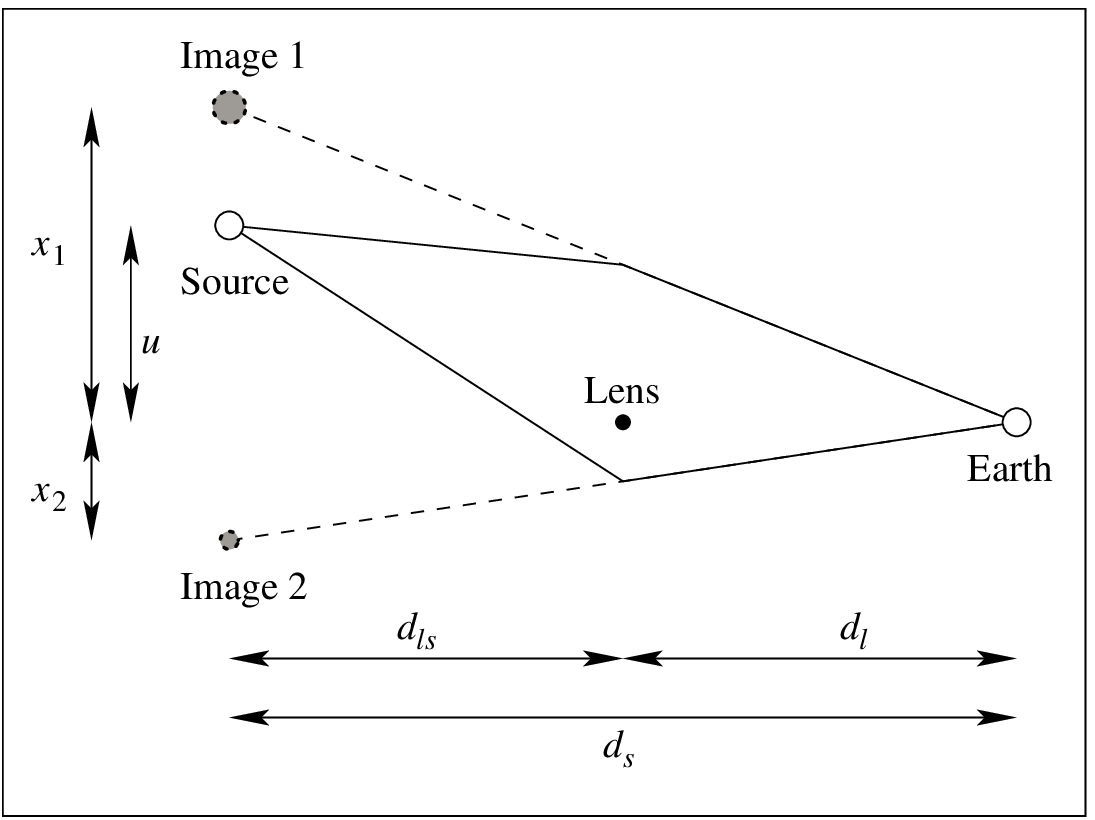}{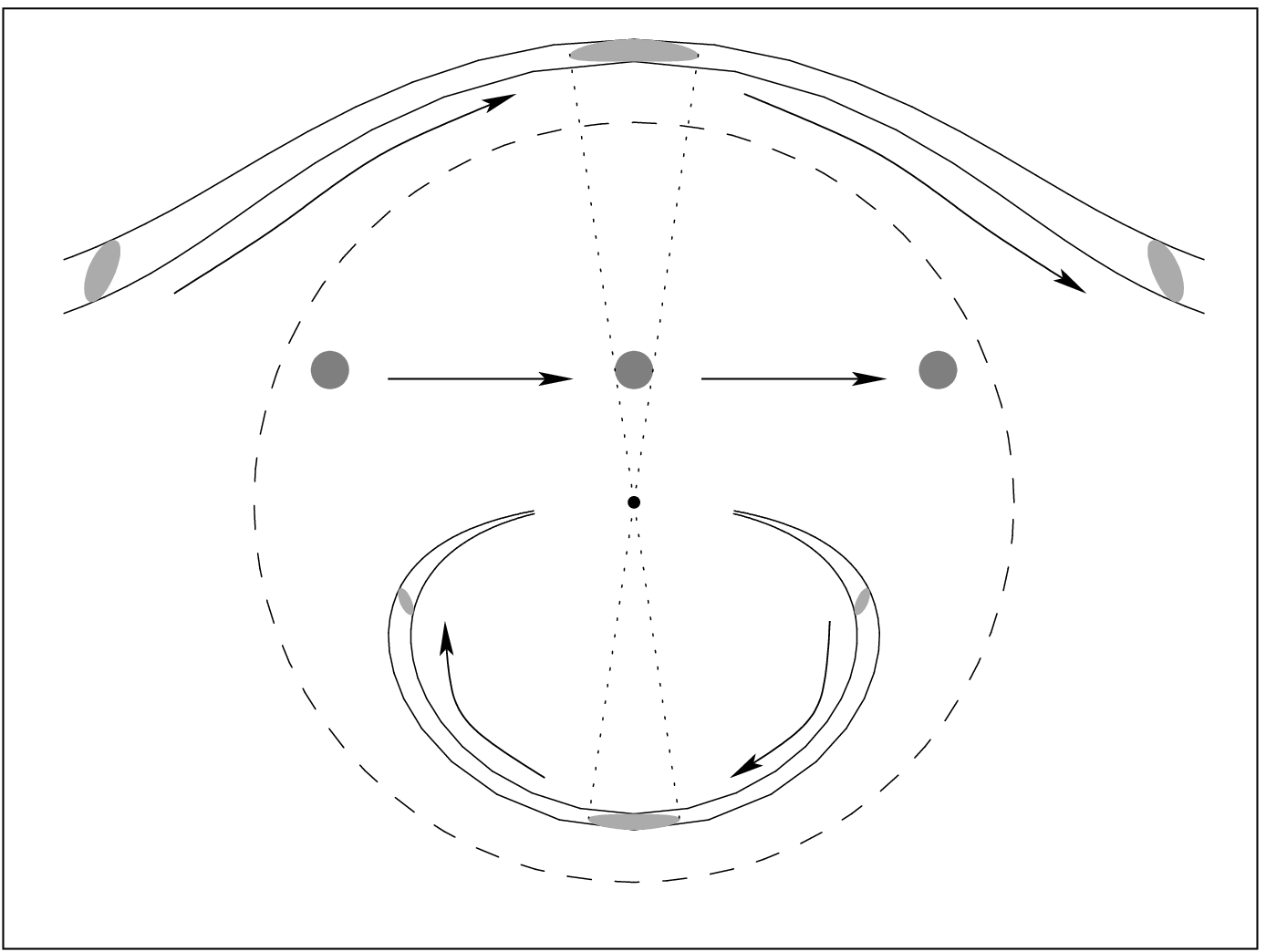}
\caption{(a) Illustration of the lensing geometry.  Two images of the
source are produced, at angular distances $x_1$ and $x_2$ from the
observer-lens line of sight.  The distances $d_l$, $d_s$, and $d_{ls}$
are between observer and lens, observer and source, and lens and
source, respectively.
(b) Illustration of the two images' motion vs. time.
The dashed circle indicates the Einstein ring. The gray circles show
the source at different times, while the light gray ellipses show the
corresponding images. 
\label{lensgeom}}
\end{figure}

In this section, we will focus on the case of a single lens and single
(unresolved) source, since this describes the vast majority of lensing
events.  This situation was completely analyzed by \citet{pac86}, and
we follow his notation here.  Consider the geometry shown in 
Fig.~\ref{lensgeom}.  This shows a source at distance $d_s$ behind a
lens at distance $d_l$.  The lens has an angular Einstein radius 
given by 
\begin{equation}
\thetaE=\sqrt{\frac{4GM}{c^2}\frac{d_{ls}}{d_ld_s}},
\label{thetaE}
\end{equation}
where $M$ is the lens mass and $d_{ls}=d_s-d_l$ is the distance
between lens and source.  If the angular impact parameter between
source and lens, in units of $\thetaE$, is $u$, then two images are
produced on the sky along the lens-source axis at positions relative
to the lens of 
\begin{equation}
x_{1,2}=\frac{\thetaE}{2}\left(u\pm\sqrt{u^2 + 4}\right).
\label{x}\end{equation}
These images are magnified relative to the intrinsic source brightness
by factors 
\begin{equation}
A_{1,2}=\frac{u^2+2}{2u\sqrt{u^2+4}}\pm\frac{1}{2}
\label{A}\end{equation} 
respectively.  The total magnification is the sum of these two terms, 
\begin{equation}
A=A_1+A_2 = \frac{u^2+2}{u\sqrt{u^2+4}}=2A_1-1 = 2A_2+1.
\end{equation}

In general, the lens and source exhibit finite relative proper motion,
meaning that the impact parameter is a function of time.  Usually, the
relative motion is well approximated as linear in time, and can be
written ${\bm u}(t) = {\bm u}_{\rm min} + {\bm\mu}t/\tE$, measuring
$t$ with respect to the time of smallest impact parameter ${\bm
u}_{\rm min}$.  Here, $\tE$ is the event timescale, the time it takes
the lens to move one angular Einstein radius relative to the
observer-source line of sight, and ${\bm\mu}$ is the unit vector on
the sky along the relative velocity.  Eqns.~\ref{x} and \ref{A} then
give $x_{1,2}(t)$ and $A_{1,2}(t)$ as a function of time.  An example
is shown in Fig.~\ref{lensgeom}.

As noted by \citet{boden}, the center-of-light of the two images
shifts relative to the unlensed source position by
\begin{equation}
\Delta{\bm\Theta}_{\rm CoL} = \thetaE \frac{{\bm u}}{u^2+2}
\label{astrometric}
\end{equation}
Also, note that at any given time, the separation between the
two images is  
\begin{equation}
\Delta s=\thetaE\sqrt{u^2+4}, \label{deltas}
\end{equation}
and the ratio of magnifications is 
\begin{equation}
R=\frac{A_2}{A_1} = \frac{A-1}{A+1}
= \frac{u^2+2-u\sqrt{u^2+4}}{u^2+2+u\sqrt{u^2+4}}. \label{ratio}
\end{equation}
These also become functions of time.

It is apparent that the magnitude of the image motion is comparable to
the angular Einstein radius.  For a bulge event with $M=0.3
M_\odot$, $d_l=4$ kpc and $d_s=8$ kpc, this is roughly $\thetaE=0.55$
mas, and scales like $M^{1/2}$.  In comparison, the resolution of the
Hubble Space Telescope is $\sim100$ mas, while the largest single
aperture telescope (the 10m Keck) has a resolution using adaptive
optics of roughly $\sim 50$ mas.  To have any hope of measuring $\thetaE$,
multi-element interferometric arrays will be required.  In the next
sections we describe how interferometers may be applied towards
microlensing.

\section{Review of Interferometry}

A stellar interferometer combines starlight from two or more
apertures; the resulting interference fringes can be used to derive a
great deal of information about the source being observed, including 
the source location (astrometry) and intensity distribution (imaging). 

\subsection{Astrometry}

For observations with a finite bandwidth a fringe pattern will be
formed when the optical pathlengths from the star, through the
instrument, to the beam combination have been equalized.  Given an aperture
separation, called the baseline $\vec{B}$, and a star in the direction
$\vec{s}$ it follows from simple geometry that an additional delay $d$
must be introduced into one of the arms of the interferometer, given
by
\begin{equation}
d = \vec{B} \cdot \vec{s} + c
\end{equation}
where $c$ is any additional static delay internal to the
interferometer.  It follows that the differential delay $\Delta d$
between two stars can be used to determine the angle $\Delta \vec{s}$
between them. Using laser metrology systems it is possible to measure
internal delay differences at the nanometer level, which for a long
baseline interferometer corresponds to astrometric precision at the
micro-arcsecond level. This is the basis for interferometric
astrometry, which has been demonstrated to achieve 100 micro-arcsecond
precision \citep{shao00}, and is expected in the next several years to
produce measurement precision approaching 10 micro-arcseconds
\citep{boden99, delplancke00}.

The practical aspects of very high precision interferometric
astrometry are beyond the scope of this paper. However, the technique
is quite challenging and requires a rather complex instrument design.
The instruments currently being designed and built are not expected to
begin operations until at least 2004-2005. In addition, due to the
effects of atmospheric seeing, the very highest astrometric precision
is likely only attainable by space-based interferometers. 

\subsection{Fringe Visibility}

The fringe pattern measured by an interferometer can be described as a
complex quantity called the visibility ($\hat{V}_\lambda$). It is
related to the source brightness distribution on the sky
($I_\lambda(\vec{s})$) via 
\begin{equation}
\hat{V} = \frac{\int I(\vec{s}) e^{- \frac{2 \pi i}{\lambda} \vec{s} \cdot \vec{B}} {\rm d} \Omega}{\int I(\vec{s}) {\rm d} \Omega}
\label{vCZ}
\end{equation}
where $\lambda$ is the wavelength of observation and $\vec{s}$ points in
the direction of the source.  This relation is known as the van
Cittert-Zernike theorem \citep{tms,lawson}. 
In effect, the source intensity distribution and the 
the fringe visibility measured by an interferometer are related via a Fourier transform.

In the case of a microlensing event, as long as the individual lensed
images are unresolved by the interferometer, $I$ can be
modeled as two point sources located at $\vec{s}_0$ and $\vec{s}_1$.
\begin{equation}
I(\vec{s}) = I_0 \delta(\vec{s}-\vec{s}_0) 
+ I_1 \delta(\vec{s}-\vec{s}_1)
\end{equation}
Defining the intensity ratio $R = I_1 / I_0$ and $\Delta\vec{s} = \vec{s}_1 - \vec{s}_0$ we find
\begin{equation}
\hat{V} = \frac{1}{1+R} \left( 1 + R\ 
e^{- \frac{2 \pi i}{\lambda} \Delta\vec{s} \cdot \vec{B}} \right)
\end{equation}

\noindent It is useful to write this visibility in a slightly different form
\begin{equation}
\bar{V} = |\hat{V} | e^{i \phi}
\end{equation}
where 
\begin{equation}
 |\hat{V}|^2 = \frac{1}{(1+R)^2}\left( 1 + R^2 + 2 R \cos(\frac{2 \pi}{\lambda} \Delta\vec{s} \cdot \vec{B}) \right)
\label{vsq}\end{equation}
The quantity $|V|^2$ (or simply $V^2$) is usually what is measured by an optical
interferometer; it corresponds to the contrast of the observed
fringes. The phase of the complex visibility is given by 
\begin{equation}
\phi = \arctan \left( \frac{R \sin(\frac{2 \pi}{\lambda} \Delta\vec{s} \cdot \vec{B})}{1 + R \cos(\frac{2 \pi}{\lambda} \Delta\vec{s} \cdot \vec{B})}\right)
\end{equation}
However, the phase measured by a single pair of apertures is corrupted
by the effects of atmospheric turbulence on very short timescales (in
the optical and infrared $\sim 1$ radian in $\sim 10$ milliseconds)
and contains no useful information, unless it is measured
simultaneously with a phase reference source --  that is in
fact identical to the astrometry approach discussed in the previous
section.

It is also important to note that $V^2$ measurements are subject to a
wide variety of error sources, and that most of these errors introduce
{\em bias} in the measurement \citep{tango80,colavita99,guyon02}. That is
to say that these error sources -- which can include instrument
vibrations, polarization mismatch, optical alignment errors, and the
blurring effects of atmospheric turbulence -- tend to {\em reduce}
the measured fringe visibility amplitude. This makes calibration of
measured visibilities quite challenging.  Modern optical
interferometers measure $V^2$ for one or a few baseline orientations;
the precision attained is typically a few percent \citep{boden98b} and in
some cases as high as 0.3 \% for bright stars under favorable conditions
\citep{coude}. Unfortunately, it is not clear that the new generation of
large-aperture interferometers equipped with AO systems and single-mode 
fiber spatial fibers will be able to attain such high visibility 
precision \citep{guyon02}.

\subsection{Closure Phase}

Despite the phase corruption introduced by the atmosphere, it is still
possible to recover limited phase information without resorting to the
technically complex phase referencing method, provided one
interferometrically combines at least three apertures. In such a case
one can form a quantity by multiplying the three complex visibilities
formed over the three baselines.  Since the visibility is the Fourier
transform of the surface brightness distribution (c.f. Eqn~\ref{vCZ}), 
the product of three visibilities measured over a closed triangle is
the surface brightness's bispectrum $B({\bm k}_1,{\bm k}_2)$,
evaluated at
${\bm k}_1=(u_1,v_1)$ and ${\bm k}_2=(u_2,v_2)$ corresponding to the
legs of the triangle.  The phase of the measured bispectrum is
known as the closure phase \citep{lawson}. The closure phase is immune
to many forms of atmospheric 
corruption, which can be illustrated as follows: above each aperture
there is a column of atmosphere with time-variable parcels of
differing indices of refraction and hence optical pathlength.  Thus
the atmosphere above each aperture contributes a time-variable phase
error, giving
\begin{equation}
\bar{V} = |\hat{V} | e^{i ( \phi_{12} + \phi_1 - \phi_2 )}
\end{equation}
where $\phi_1$ and $\phi_2$ are the phase errors associated with 
apertures 1 and 2 respectively, and $\phi_{12}$ is an intrinsic 
phase associated with the source as measured by the 1-2 baseline.

The bispectrum is thus
\begin{eqnarray}
\bar{V}_{123}& = &|\hat{V}_1 ||\hat{V}_2 ||\hat{V}_3 | e^{i ( \phi_{12} + \phi_{23} + \phi_{31} + (\phi_1 - \phi_2) + (\phi_2 - \phi_3) + (\phi_3 - \phi_1) )} \nonumber \\
	     & = & |\hat{V}_1 ||\hat{V}_2 ||\hat{V}_3 | e^{i (\phi_{12} + \phi_{23} + \phi_{31}) }
\end{eqnarray}
We see that the atmospheric phase errors (as well as many other
aperture-dependent phase errors) cancel. This is a well-known result,
first applied in radio interferometry \citep{jennison}. 
However, it is not immediately obvious what the closure phase
represents. Below we derive an expression relating the observed
closure phase to the binary point source representing a microlensing
event.

Assume 3 apertures, resulting in 3 baselines $\vec{B}_1, \vec{B}_2$
and $\vec{B}_3$. Note that 
\begin{equation}
\vec{B}_1 + \vec{B}_2 + \vec{B}_3 = 0. 
\label{zerosum}\end{equation}
As before, we are looking at two point sources with intensity ratio
$R$ and separation $\Delta\vec{s}$. On each baseline we measure
a visibility $\hat{V}_n$ given by Equation~(\ref{vCZ}).
\begin{eqnarray}
V_{123} & = & \hat{V}_1 \hat{V}_2 \hat{V}_3 \nonumber \\
   & = & \frac{1}{(1+R)^3}\left( 1 + R e^{- \frac{2 \pi i}{\lambda} \Delta\vec{s} \cdot \vec{B}_1} \right)\left( 1 + R e^{- \frac{2 \pi i}{\lambda} \Delta\vec{s} \cdot \vec{B}_2} \right)\left( 1 + R e^{- \frac{2 \pi i}{\lambda} \Delta\vec{s} \cdot \vec{B}_3} \right) \nonumber \\
	& = & \frac{1}{(1+R)^3} [ 1 + R(e^{- \frac{2 \pi i}{\lambda} \Delta\vec{s} \cdot \vec{B}_1} + e^{- \frac{2 \pi i}{\lambda} \Delta\vec{s} \cdot \vec{B}_2} + e^{- \frac{2 \pi i}{\lambda} \Delta\vec{s} \cdot \vec{B}_3} ) \nonumber\\
 & & + R^2(e^{ \frac{2 \pi i}{\lambda} \Delta\vec{s} \cdot \vec{B}_1} + e^{ \frac{2 \pi i}{\lambda} \Delta\vec{s} \cdot \vec{B}_2} + e^{ \frac{2 \pi i}{\lambda} \Delta\vec{s} \cdot \vec{B}_3} )  + R^3 ]
\end{eqnarray}
the closure phase is thus
\begin{equation}
\phi_{123} = \arctan \left( \frac{ (R^2 - R)\sum_{i=1,2,3}{\sin(\frac{2 \pi}{\lambda} \Delta\vec{s} \cdot \vec{B}_i)} }{ 1 + R^3 + (R + R^2)\sum_{i=1,2,3}{\cos(\frac{2 \pi}{\lambda} \Delta\vec{s} \cdot \vec{B}_i)}   }  \right)
\label{phi123}
\end{equation}

There are a few things to note: the closure phase is always zero when
$R=1$, i.e. the source is symmetric. In addition, by Taylor expanding the 
sine terms and recalling Eqn.~(\ref{zerosum}) it is
easy to show that for the case when $\Delta\vec{s} \ll \frac{\lambda}{|B|}$
\begin{equation}
\phi_{123} \propto \left( \frac{ \Delta\vec{s} }{\lambda / |B| } \right)^3
\label{cubescale}
\end{equation}
This implies that a source must be resolved by the interferometer 
in order to produce a non-zero closure phase; in the case of a 
partially resolved source the magnitude of the closure phase signal 
is very sensitive to the separation of the source components.

Unlike $V^2$, the phase measured by an optical interferometer is
largely unbiased by measurement noise. In other words, the phase noise is
zero-mean, and can be reduced by averaging over a sufficiently long
time.  The SNR for closure phase is given by
\citep{tango80,shaocola92} 
\begin{equation}
{\rm SNR}_\phi=\left[3\left(\frac{2}{NV^2}\right)
+6\left(\frac{2}{NV^2}\right)^2+4\left(\frac{2}{NV^2}\right)^3
\right]^{-1/2}
\end{equation}
as compared to that for visibility and phase,
\begin{equation}
{\rm SNR}_V=\sqrt{\frac{NV^2}{1+(NV^2)^{-1}}}
\end{equation}
In the photon-noise dominated regime ($NV^2\gg1$), the
SNR for closure phase and visibility scales as $N^{1/2}$.  However,
for photon-starved sources, the SNR drops precipitously as $N^{3/2}$,
even worse than the $\propto N$ scaling of the visibility SNR.
Hence it is important to check whether microlensing
sources will be photon-rich or photon-starved. 

In the $K$-band ($2.2 \mu$m) a 15th magnitude source produces 0.16
photons ${\rm cm}^{-2}~{\rm s}^{-1}$. Thus we would expect an 8-m
class telescope to collect $\sim 80,000$ photons sec$^{-1}$.  Assuming
a coherence time $\tau_0=20$msec, and a typical photon efficiency of
5\%, we have $N=80\gg1$ per integration, safely in the photon rich
regime when using modern low-noise detectors.  In the absence of
systematic errors a 3-element interferometer should achieve a closure
phase precision of $10^{-3}$ radians in 250 seconds. However, it
remains to be seen if such levels can be achieved. To date, closure
phase measurements have been made in the optical and near-IR by 2
groups. COAST achieves a closure phase precision of $\sim 2$ degrees
\citep{baldwin96}, while NPOI achieves phase drifts of $\sim 10$
degrees hr$^{-1}$, which can be calibrated to the level of 1-4 degrees
by looking at known single stars \citep{hummel98}. Recent progress in
integrated optics -- which due to their compact design are much less
susceptible to systematic closure phase errors -- may well yield a
large improvement in the achievable accuracy \citep{berger01}.

We suggest that a good method to correct for systematic closure phase
errors would be to measure closure phase across a range of
wavelengths. Given that many systematic error mechanisms give errors
with a characteristic wavelength dependence that varies from that of
the observed source, it should be possible to separate the signals.
To illustrate, consider the case of a physical pathlength mismatch in
the interferometric beam combiner, applied only to the light from one
pair of apertures. Such a path mismatch will give a closure phase
error that is inversely proportional to wavelength. Now consider the
typical microlensing event; although the lens is achromatic, from
equation~(\ref{cubescale}) 
we see that the amplitude of the closure phase signal is 
inversely proportional to the cube of the wavelength. Hence, measuring
the closure phase at several wavelengths should allow one to solve for
both the source closure phase and the internal closure phase error.
Clearly, this type of experiment would benefit from long baselines and
wide bandwidth; the VLTI (200-m baseline) and AMBER (3-aperture
combination from 1-2.4$\mu$m) provide such a suitable combination
\citep{glind}.

\section{Studying microlensing with interferometers}

Microlensing events are amenable to study by interferometers by all
three methods described in the previous section.  Microlensing causes
an astrometric excursion of the center of light of the two images
given by Eqn.~(\ref{astrometric}), which may be measured using narrow
angle astrometry \citep{pac98,boden}.  Similarly, as the secondary
image brightens, the two images become distinguishable from a single
point source, causing a decrement in the visibility amplitude
\citep{delplancke}.  Lastly, during the microlensing event the
distribution of flux about the center of light becomes asymmetric,
giving a signal in the closure phase.  To measure $\thetaE$ by
astrometry, multiple measurements are required to map out the
microlensing ellipse, while visibility and closure phase can give
$\thetaE$ from a single measurement.

\subsection{Expected instrumental performance}

Now let us consider the measurement of these signals in practice.  An
important point to recall is that typical microlensed sources, by
interferometrists' standards, are faint.  This has several
ramifications for the measurement of the signals in astrometry,
visibility, and closure phase. As the application of these techniques
to optical/IR wavelengths is still a relatively new field, there is
considerable uncertainty regarding the measurement precision that will
be achievable. Thus we will define a range of likely precisions (for
the optimist and the pessimist) and evaluate the usefulness of the
technique for each case.

Astrometric measurements of faint sources require the presence of a
bright ($K<13$) phase reference within an isoplanatic patch, typically
of order 20 arcseconds in the $K$ band.  The probability of a $K<13$
star falling within $20\arcsec$ is roughly 17\%
\citep{shaocolavita92b}.  Narrow-angle astrometry at the 100
micro-arcsecond level has already been demonstrated at PTI, and it is
reasonable to expect the Keck and VLTI to produce the specified
precision of 10-30 microarcseconds. However, narrow-angle
interferometric astrometry is the least mature of the three
techniques, and hence we will consider the utility of the technique
for precisions of both 100 and 10 $\mu$as.

Next consider measurement of the visibility amplitude for faint
sources. The precision with which $|V|$ can be measured with small
apertures (single-$r_0$, where $r_0$ is the Fried parameter) can be as
good as 0.3\%. However, recall that in our estimate for the number of
photons per coherence time, we assumed an aperture of 8m.  Since a
typical Fried parameter for $K$ band is $r_0\sim60$ cm, adaptive
optics (AO) are required to correct distortions in the wavefront.
Ideally, an AO system would completely flatten the wavefront, giving
uniform path delays across the aperture.  This would restore the point
spread function (PSF) to the familiar Airy pattern.  Unfortunately, AO
systems are not ideal, and in practice manage to get $\sim50\%$ of the
light into a tight core, and the remainder in a diffuse halo.  This
Strehl ratio sets the upper limit to the (squared) visibilities an
interferometer can achieve, unless a spatial filter such as a
single-mode optical fiber is used. However, such a spatial filter
increases the maximum achievable visibility amplitude at the expense
of losing light (max coupling efficiency $\sim78\%$, \citet{shak88})
and introducing visibility biases which depend on the Strehl ratio
\citep{guyon02}. In principle, these effects may be calibrated by
observing point sources and monitoring the AO Strehl ratio in real
time.  In practice, such calibration is difficult and as yet unproven.
At present, it is not known how well visibility amplitudes may be
calibrated for large, AO-corrected apertures, but it may potentially
be on the order of 5\%. In summary, the expected precision in
visibility amplitude ($|V|$) for the Keck and VLTI will likely be between 0.5
and 5\%; we will consider the resulting measurement SNR for both
cases.

Lastly, we consider closure phase.  As mentioned earlier, closure
phase is free of many of the error sources afflicting visibility
amplitude and phase.  However, other systematics can arise which we
have not anticipated.  The current generation of interferometers have
demonstrated closure phase precision at roughly the 1-degree
level. However, there is no obvious reason why performance at the
0.001 radian (or 0.06 degree) could not be achieved, in particular
since the necessary high-precision phase-measurement equipment is
being developed for differential-phase detections of hot exoplanets
\citep{akeson02}.  In addition, \citet{seg02} has discussed
the possibility of detecting hot exoplanets using differential closure
phase at the VLTI, an application which would require closure phase
measurement precision on the order of $10^{-4}$ radians.  Hence we will
consider the closure phase SNR for both pessimistic (1 deg) and
optimistic (0.001 rad) cases.

\subsection{Expected signal}

One of the differences between the study of microlensing via
astrometry, as opposed to visibility or closure phase, is that the former
does not rely upon resolving the two lensed images from each other.
Because of this, the signal itself is independent of wavelength.
However, a single astrometric measurement alone does not suffice to
measure $\thetaE$.  In principle, at least 3 astrometric measurements
are required to measure the microlensing signal and disentangle it
from proper motion of the source.  Since visibility and closure phase
do involve 
resolving the two images, a single measurement does suffice to give
$\thetaE$, and the signal is a function of $\thetaE B/\lambda$.

\begin{figure}
\plottwo{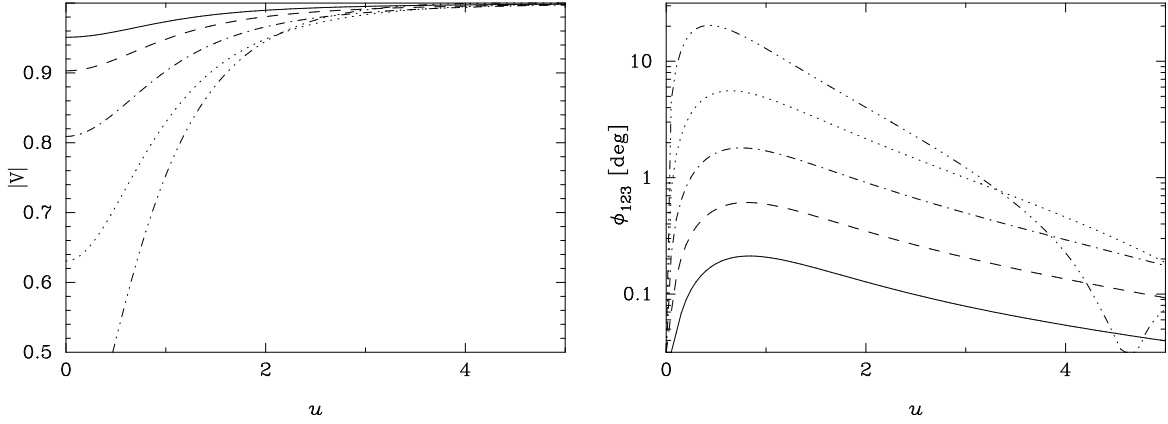}{f2b.eps}
\caption{The first panel shows the visibility amplitude as a function
of $u$ for varying degrees of resolution: $\thetaE
B/\lambda=0.05,0.07,0.1,0.14,0.2$ from top to bottom.  The  
second panel shows the corresponding closure phase, in degrees.
Note that for highly resolved events, the signals do not follow the
approximate Eqns.~(\ref{microvsq}) and (\ref{microphi}).
\label{signal}}
\end{figure}

\begin{figure}
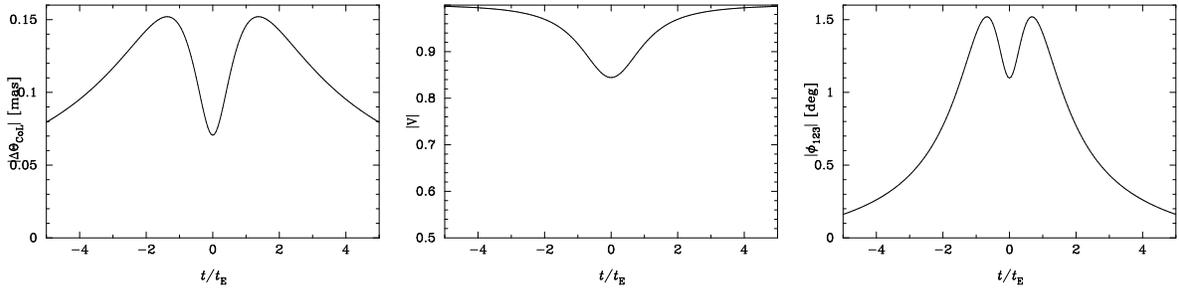

\centerline{
\includegraphics[width=0.3\textwidth]{f3a.eps}\hfil
\includegraphics[width=0.3\textwidth]{f3b.eps}\hfil
\includegraphics[width=0.3\textwidth]{f3c.eps}
}
\caption{Interferometric signal for a microlensing event with
$\thetaE=0.43$ mas and $\umin=0.35$.  The first panel shows the
magnitude of the astrometric motion of the center of light.  The
second panel shows the visibility amplitude, and the third panel plots
the closure phase.  We have assumed a three element
interferometer in an equilateral triangle, with arms of 100 m each,
observing at 2.2 $\mu$m.  We also assume that Earth rotation allows
a maximal projection of the image separation along at least one of the
baselines.
\label{example}}
\end{figure}

Inserting Eqns.~(\ref{deltas}) and (\ref{ratio}) into Eqn.~(\ref{vsq})
and expanding to lowest order in $\theta_E$, we see that the
visibility signal behaves as
\begin{equation}
1-V^2\approx\left(2\pi\frac{B}{\lambda}\thetaE\right)^2
\frac{u^2+4}{(u^2+2)^2}
\label{microvsq}
\end{equation}
for $\thetaE\ll\lambda/2\pi B$.  For a 3-element interferometer in an
equilateral triangle, inserting Eqns.~(\ref{deltas}) and
(\ref{ratio}) into Eqn.~(\ref{phi123}) gives to lowest order
\begin{equation}
\tan(\phi_{123})\approx\left(\pi\frac{B}{\lambda}\thetaE\right)^3
\frac{u\left(u^2+4\right)^2}{\left(u^2+2\right)^3}
\label{microphi}
\end{equation}
again for $\thetaE\ll\lambda/\pi B$.
Figures~\ref{signal}, ~\ref{example} and ~\ref{SNR} illustrate this.
The steep dependence on the resolution means 
that there is a strong incentive to
measure closure phase at the smallest possible wavelength.  For
example, the closure phase in $J$ band ($1\mu$m) is more than 10 times
larger than that in $K$ band (2.2 $\mu$m).

An actual microlensing event will produce time-varying signals in
astrometry, visibility and closure phase, analogous to the familiar
time-varying photometric lightcurves. Consider a typical microlensing
event with a $0.5 M_\odot$ lens at distance $d_l=6$ kpc, giving
$\thetaE=0.43$ mas.  Assume $\umin=0.35$, so that the maximum
magnification is $A=3$, and the maximal brightness ratio of the images
is $R=1/2$.  An interferometer with a 100m baseline operating at 2.2
$\mu$m ($K$-band) has a resolution of 4.5 mas.  If we have three such
baselines arranged in an equilateral triangle, they would observe the
signals in astrometry, visibility, and closure phase shown in
Fig.~\ref{example}.  The astrometric signal has a maximal amplitude of
$\sim0.35\thetaE$, which in this example is about 0.15 mas.  The
visibility, which is usually minimized at closest approach
($u=\umin$), is here $|V|=0.85$.  The closure phase has a maximum
value of $\phi_{123}=1.5^\circ$.  With present technology, precisions
of $\sim 5\%$ in $|V|$ and $\sim1^\circ$ in closure phase may be
achieved, so at present such events should be studied by visibility.
However, advances in measurement precision and observation at shorter
wavelengths can significantly enhance the signal-to-noise.  In the
next subsection, we consider a reasonable range for the expected
instrumental performance and show how this affects the prospects for
determination of $\thetaE$ using interferometers.

\subsection{Prospects for mass measurement}

\begin{figure}
\plottwo{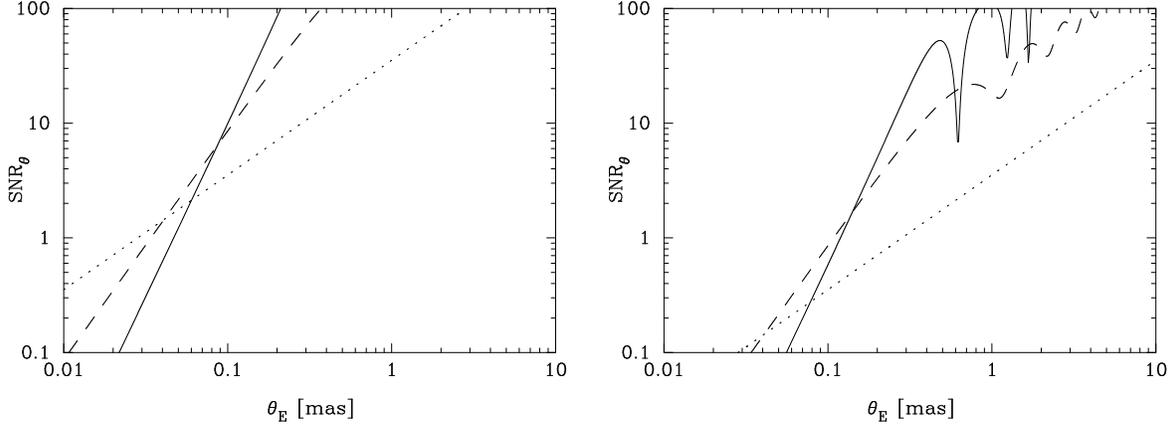}{f4b.eps}
\caption{SNR in $\thetaE$.  The solid line corresponds to the
signal-to-noise ratio in $\thetaE$ derived from closure phase, the
dashed line corresponds to visibility, and the dotted line corresponds
to astrometry. The left panel shows `optimistic' signal-to-noise
ratios, assuming 10 $\mu$as errors in the astrometry, 
0.5\% errors in $|V|$ and 0.001 rad errors in the
closure phase.  The right panel shows `pessimistic' signal to noise
using 100 $\mu$as astrometric errors, 
5\% errors in  $|V|$ and $1^\circ$ errors in closure phase.  We
assume astrometric measurements are taken at $u=1.4$, 
closure phase is measured at $u=0.86$ in $J$ band (1 $\mu$m)
and visibility is measured at $u=0.35$ in $K$ band (2.2 $\mu$m) where
higher Strehl ratios may be achieved.
\label{SNR}}
\end{figure}

Except for highly resolved events, the astrometric signal scales as
$\sim\thetaE$, while the visibility signal scales as $\sim\thetaE^2$
and the closure phase as $\sim\thetaE^3$.  These steep scalings
improve the SNR in the derived Einstein radius; for example a SNR of 5
in visibility translates into a SNR of 10 in $\thetaE$, while a
closure phase SNR of 5 becomes a $\thetaE$ SNR of 15.  
Figure~\ref{SNR} plots the SNR in
$\thetaE$ from astrometry, visibility and closure phase, assuming
respectively optimistic errors (left panel) of $10\mu$m in astrometry,
0.5\% in $|V|$, and 0.001 rad in closure phase, and (right panel)
pessimistic errors of 100 $\mu$m astrometry, 5\% visibility amplitude,
and $1^\circ$ closure phase.  
If errors of 0.001 can be achieved in $J$ band, then events with 
$\thetaE\gtrsim0.1$ mas can be measured with SNR$\gtrsim 10$.  

\begin{figure}
\plottwo{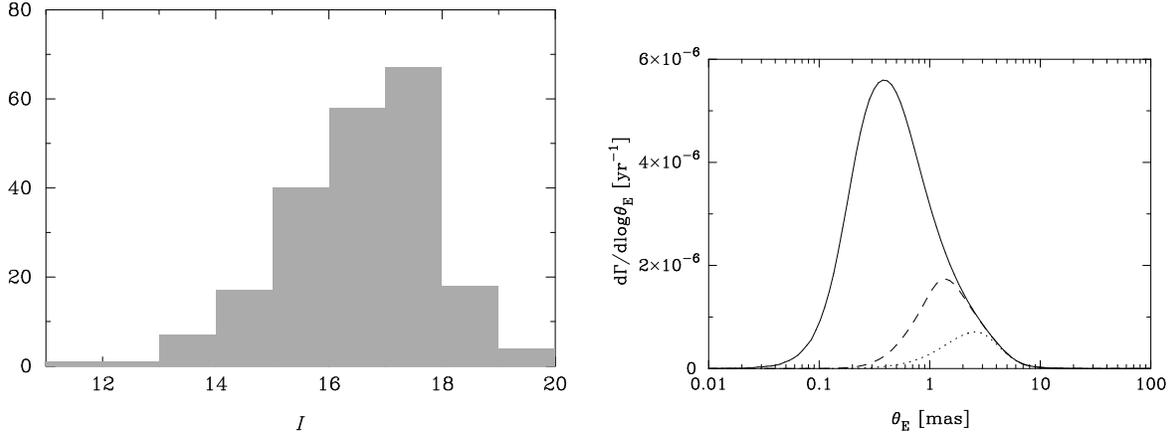}{f5b.eps}
\caption{The first panel shows a histogram of $I$ magnitudes for 213
microlensing events observed by OGLE during 1997-1999.  Of these, 26
have $I<15$, and 48 have $I<15.5$.  The second panel shows the
distribution of lensing rate with respect to angular Einstein radius,
$d\Gamma/d\log\thetaE$.  The solid line shows the total rate
distribution, the dashed line shows the distribution of events with
$\tE>50$ days, and the dotted line shows that for events with
$\tE>100$ days.
\label{events}}
\end{figure}

The next step is to determine the brightnesses and Einstein radii for
typical events.  Figure~\ref{events} illustrates the properties of
bulge events.  The first panel shows the distribution of peak $I$
magnitudes (or in cases where the peak was not observed, the brightest
observed magnitude) for events observed by the OGLE collaboration
during the 1997-1999 seasons.  Assuming that sources with $K<14$ can
be observed interferometrically, and assuming that $I-K=1.5$, we
estimate that roughly 20\% of events are bright enough to be followed
up with interferometers.  However, future surveys may go deeper than
OGLE-II, returning a smaller fraction of bright events.  
Next, we turn to the distribution of
Einstein radii.  Since this has not been measured, we must estimate
theoretically what the distribution will be.

The rate of microlensing events depends upon the properties of the
lenses and source stars. We assume two types of stars, bulge stars
and disk stars.  
For the disk stars, we use the mass function of \citet{gbf97},
namely $dN/dM\propto (M/M_b)^{\alpha}$, with $M_b=0.59 M_\odot$, and
$\alpha=-0.56$ for $M_{\rm min}<M<M_b$, and $\alpha=-2.21$ for
$M_b<M<M_{\rm max}$.  We take $M_{\rm min}=0.1M_\odot$ and $M_{\rm
max}=100M_\odot$.  The disk density profile is taken to be
$\rho(r_l)=\rho_0 e^{r_l/r_d}$, where $\rho_0=8~10^7 M_\odot/{\rm
kpc}^3$ is the local density, $r_d=3.5$ kpc is the disk scale length,
and $r_l$ again is the distance of the lens from us towards the
source, taken to be at the Galactic center.  The velocity distribution
of disk stars is taken to be a flat rotation
curve $v_c=220$ km/s, along with velocity dispersion in each
transverse direction of $\sigma_d=30$ km/s.  For the bulge stars, we
use the mass function of \citet{zoccali}, which has the same form as
the disk MF but with $M_b=1M_\odot$, $\alpha=-1.33$ for
$0.15M_\odot<M<M_b$, and $\alpha=-2$ for $M_b<M<100M_\odot$.  The
bulge density distribution we use is based on the barred model of
\citet{hangould}, with
$\rho(r_l)=\rho_0\exp[-\frac{w}{2}(r_s-r_l)^2]$, with central density
$\rho_0=2.07~10^9 M_\odot/{\rm kpc}^3$, $w=(\cos i/x_0)^2+(\sin
i/y_0)^2$ with scale lengths $x_0=1.58$ kpc, $y_0=0.62$ kpc and
inclination $i=20^\circ$.  Bulge stars are taken to have no net rotation
and a velocity dispersion in each transverse direction of
$\sigma_b=110$ km/s.

Lenses are drawn from both the disk and bulge populations.  
For simplicity, we assume that all sources lie at the galactic center,
with the bulge velocity distribution.  We neglect the motion of the
local standard of rest relative to the flat rotation at $v_c$.  
The distribution of Einstein radii
is given in the Appendix by Eqn.~(\ref{dGdthE}) and
Eqn.~(\ref{vtran}). Using the above parameter values, the resulting
rate distribution is 
plotted in the second panel of Fig.~\ref{events}.  Clearly, a large
fraction of events should have $\thetaE$ accessible to instruments like
the Keck Interferometer or VLTI.  

As mentioned earlier, for very long timescale events ($\tE>50$ days)
the parallax may be measured with sufficiently high quality
photometry, meaning that measurement of $\thetaE$ gives the mass and
distance.  Estimating that $\sim 15\%$ of events are of sufficiently
long duration, that $\sim10\%$ are bright enough to be observed with
VLTI or Keck, and that 1000 events are detected every year 
we expect to measure masses for about 15
microlensing events from the ground every year.  This should be
interesting for constraining the high end of the mass function. 

\begin{figure}
%\plotone{fig6.eps}
\includegraphics[width=0.5\textwidth]{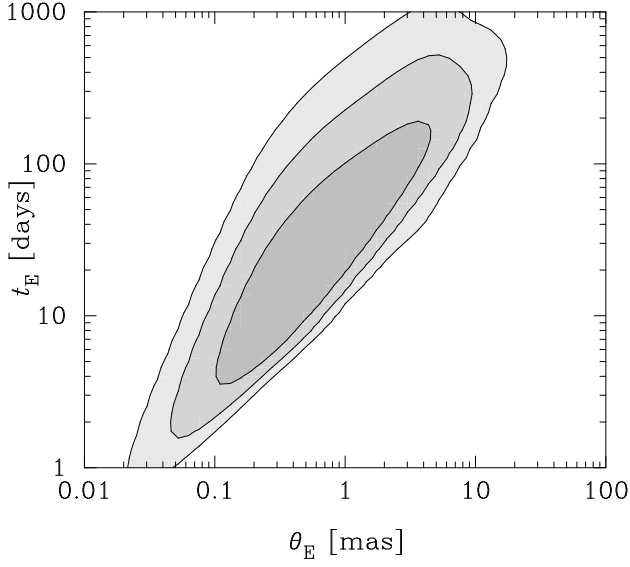}
\caption{Contours of $d\Gamma/d\log\thetaE d\log\tE$, spaced by decades.
\label{dgdthdt}}
\end{figure}

However, VLTI and Keck can observe many more events than just
those showing parallax, if there is justification to do so -- for
example, if such measurements could better constrain the lens
properties.  Let us wildly speculate about a massive campaign from
VLTI to observe 100 events per year.  
The basic observable from such a campaign is the joint distribution of
event timescales $\tE$ and angular Einstein radii $\thetaE$.  An
expression for $d\Gamma/d\thetaE d\tE$
is derived in the appendix in Eqn.~(\ref{dGdthEdtE}).
The resulting rate distribution is plotted in Fig.~(\ref{dgdthdt}).

A massive survey can measure a noisy sample of this distribution.  To
estimate how well these measurements may be used to constrain the
underlying lens properties, we use the Fisher matrix
\citep{gould_mvb,tegmark}.  For a likelihood distribution $L$ (here
$d\Gamma/d\thetaE d\tE$) with parameters $\lambda_i$, the covariance
matrix of the parameters estimated by sampling $L$ is given by
\begin{equation}
c_{ij} = ({\bf b}^{-1}){ij}
\end{equation}
where the matrix {\bf b} has elements\footnote{We follow
\citet{gould_mvb}'s notation for the Fisher matrix, which differs from
\citet{tegmark}'s notation by a partial integration.}
\begin{equation}
b_{ij}=\left\langle\frac{\partial\log L}{\partial\lambda_i}
\frac{\partial\log L}{\partial\lambda_j}\right\rangle
\end{equation}
and the angular brackets denote integration over possible observables
(here $\thetaE$ and $\tE$) weighted by $L$.
We vary eight parameters: the bulge
velocity dispersion $\sigma_b$, mass function parameters $\alpha_1$,
$\alpha_2$ and $m_p$, and the same parameters for the disk lenses.  
Results are listed in Table~\ref{fisher}.  
We find that the corresponding disk and bulge parameters are somewhat
degenerate, as may be expected, but that large numbers of events can
allow constraints to be placed on the bulge parameters.  For example,
with $N=400$ lenses the error on the slope of the high end of the
bulge mass function becomes $\Delta\alpha_2=0.2$, and the error on
$m_{p,b}$ becomes $0.18 M_\odot$, sufficient to detect a break in the
mass function and to localize the break position.  The caveat to this
statement is that we have made several simplifying approximations,
for example not exploring degeneracies between the mass function
parameters and the density profile parameters.  

\begin{table}
\begin{tabular}{ccc}
parameter&value&error ($N=1$)\\
\hline
$\sigma_b$ & 110 km/s & 108 km/s \\
$m_{p,b}$ & 1 $M_\odot$ & 3.6 $M_\odot$ \\
$\alpha_{1b}$ & -1.33 & 7.9 \\
$\alpha_{2b}$ & -2 & 3.8 \\
$\sigma_d$ & 30 km/s & 466 km/s \\
$m_{p,d}$ & 0.59 $M_\odot$ & 4.0 $M_\odot$ \\
$\alpha_{1d}$ & -0.56 & 43.3 \\
$\alpha_{2d}$ & -2.21 & 9.3 \\
\end{tabular}
\caption{Errors on parameters from measuring $d\Gamma/d\thetaE d\tE$.
The second column lists the adopted values for each parameter, and the
third column lists the measurement errors estimated using the Fisher
matrix.  The errors values are to be multiplied by $N^{-1/2}$, where
$N$ is the number of bright lens events observed by the interferometer.
\label{fisher}}
\end{table}

\section{Discussion}

Interferometry is poised to revolutionize the study of microlensing
events.  Until now, microlensing has suffered the difficulty that
masses of individual lenses cannot be measured, severely limiting the
information able to be extracted from lensing surveys.  The
problem has been that the two quantities needed to measure mass and
distance, the relative parallax $\piE$ and the angular Einstein radius
$\thetaE$, are not regularly measured.  The parallax may be measured
for long duration events with high quality photometry, however
measurement of $\thetaE$ requires resolution on the order of a
milliarcsecond, necessitating interferometers.
The upcoming Space Interferometry Mission (SIM) should
measure masses and distances for a large sample of lenses, answering
the question of the microlenses' nature.  Well in advance of SIM,
however, ground-based interferometers can also provide useful
measurements of lensing events.

As mentioned earlier, microlensed sources are generally much fainter
than the typical sources studied by optical interferometers, meaning
that large apertures ($\sim 8$m) are required.  Both the Keck
Interferometer and the VLTI can measure visibility amplitude for
microlensing events using their largest apertures, but only VLTI can
measure closure phase using 3 large apertures; the Keck would be
required to employ one of the 1.8 m outrigger telescopes which collect
considerably fewer photons.  The signal measured by interferometers is
a function of the Einstein radius in units of the resolution, $\thetaE
B/\lambda$.  Because of this, there is great advantage to go to
shorter wavelengths.  However, shorter wavelengths require the use of
adaptive optics (AO) systems.  Since AO makes the accurate calibration
of visibility very difficult, but has a smaller impact on the
calibration of closure phase, there are obvious advantages to using
closure phase.  For a numerical example, a microlensing event with
$\thetaE=0.5$ mas could be observed at 10 $\mu$m without AO, but
$\thetaE B/\lambda\sim0.02$ giving a visibility signal of
$V\approx0.99$, which would be extremely challenging to distinguish
from a point source.  The same event observed at 2.2 $\mu$m using AO
has $\thetaE B/\lambda\sim0.11$ giving a closure phase signal
$\phi_{123}\approx 2.5^\circ$ which can already be measured.

Only a small fraction of events are expected to be
bright enough ($K\lesssim 14$) to be observed interferometrically.
However, certain fainter events may also be accessible to
interferometers.  If a bright ($K<13$) star falls within the
isoplanatic angle, then phase referencing may be employed to extend
the coherence time significantly.  As noted earlier, for sites with
small isoplanatic patches the probability of finding a suitable
bright star is poor.  Additionally, this technique is quite complex and
as yet unproven, however in principle this could allow the study of
microlensing events as faint as 20th magnitude, reaching the bright
end of LMC events.  Phase referencing must be employed to perform
narrow angle astrometry; our results indicate that events for
which phase referencing is possible may be more profitably studied
with visibility or closure phase.

We expect that $\sim15$ events every year will be bright enough
($K<14$) and have sufficiently long duration ($\tE>50$ days) to permit
the measurement of mass and distance.  We have shown that a fairly
large fraction of events accessible to ground-based interferometers
should allow measurement of $\thetaE$ with high signal to noise.  
We also investigated the prospects for a massive follow-up campaign by
VLTI, and found that statistical information on the $\thetaE$
distribution, even without individual mass measurements, can allow
constraints to be placed on lens properties like their mass function.

Even if our estimates turn out to be overly optimistic, interferometers
will still be able to elucidate the nature of claimed
black hole candidates \citep[e.g.][]{mao,bennett}.
\citet{agol} have suggested that current microlensing data indicate
the presence of a significant population of intermediate mass black
holes roaming the Galactic disk; interferometers will
be able to confirm or reject this possibility.

In this paper, we have focused on ground-based interferometers,
however our results apply also to space-based interferometers like the
Space Interferometry Mission (SIM).  SIM is primarily an astrometric
instrument, however it can also measure fringe visibilities.
Nominally, the target precision expected for SIM is 1\% in $V^2$
(M.~Shao 2002, priv.\ comm.).
SIM's baseline is 10m, and typical wavelengths are
$\lambda\approx0.6\mu$m, giving resolution of about 12 mas.  Hence, SIM
can determine $\thetaE$ with SNR$>10$ from visibility alone, entirely
independently of the astrometric determination, for events with
$\thetaE>0.44$ mas.  From Figure~\ref{events} we see that this
comprises a large fraction of the events.
The visibility measurements come for free with
the astrometric measurements, and should significantly increase
the precision of SIM mass measurements, as long as effects such as
crowding do not pose too great an obstacle.  In addition to measuring
$\thetaE$, SIM also determines $\piE$, the lens parallax, by measuring
the time of the peak of the photometric lightcurve.  Since the peak of
the visibility signal coincides with the peak of the photometric
signal, SIM visibility measurements could also be used to determine
$\piE$.  However, since the variation in $1-V^2$ is so shallow near
the peak this method may not prove to be as precise as ordinary
photometric parallax measurement.

One of the most (potentially) exciting prospects 
is a topic we have not discussed in this paper, binary microlensing.
For a binary lens system, complicated caustic structures can arise,
leading in favorable cases to the production of 5 images of the
source.  For a spectacular example of this, see \citet{an}.  During
caustic crossings, the magnification can get exceptionally large,
e.g. factors of 30, making these events bright enough to observe with
interferometers.  The five
images are currently unresolvable from each other, however VLTI and
Keck offer the prospect of imaging the multi-image pattern.  With 
a multi-aperture system (required for closure phase), one obtains
several visibility measurements 
and one closure phase at the same time, possibly allowing the
reconstruction of complex events such as caustic crossings.

\bigskip
\acknowledgments{We thank the Harvard-Smithsonian Center for
Astrophysics and the Michelson Interferometry Summer School for
hospitality while this work was carried out.  We also thank M. Mark
Colavita and Kim Griest for a careful reading of the manuscript and
helpful suggestions.  This work was supported
by JPL contract 1226901 and by the Department of Energy under grant
DOE-FG03-97ER40546. B.F.L gratefully acknowledges the support of NASA
through the Michelson fellowship program.  N.D. gratefully
acknowledges the support of NASA through Hubble Fellowship grant
\#HST-HF-01148.01-A awarded by the Space Telescope Science Institute,
which is operated by the Association of Universities for 
Research in Astronomy, Inc., for NASA, under contract NAS 5-26555.}

\appendix
\section{Appendix - Microlensing rate}
The microlensing optical depth $\tau=\int n\sigma dl$ is given by
\begin{equation}
\tau=\int_{M_{\rm min}}^{M_{\rm max}} dM\frac{dN}{dM}
\int_0^{r_s}dr_l~n(r_l)\pi r_l^2 \thetaE^2,
%\tau=\frac{4\pi G}{c^2}\int_{M_{\rm min}}^{M_{\rm max}} dM\frac{dN}{dM}M
%\int_0^{r_s}dr_l~n(r_l)\frac{(r_s-r_l)r_l}{r_s},
\label{tau}
\end{equation}
where $dN/dM$ is the mass function of lenses, normalized by
$\int(dN/dM)dM=1$, $n(r_l)=\rho(r_l)/\bar{M}$ is the number
density of lenses, $\bar{M}=\int(dN/dM)MdM$ is the average mass
and $\rho$ is the mass density of lenses.
From Eqn.~\ref{thetaE}, we then have
\begin{eqnarray}
\frac{d\tau}{d\thetaE}&=&
\int_{M_{\rm min}}^{M_{\rm max}} dM\frac{dN}{dM}
\int_0^{r_s}dr_l~n(r_l)\pi r_l^2 \thetaE^2
\delta\left[\thetaE-\left(\frac{4GM}{c^2}\frac{r_s-r_l}{r_sr_l}\right)^{1/2}
\right]\nonumber\\
&=&\frac{\pi c^2}{2G}\thetaE^3\int_0^{r_s}dr_l
\frac{r_sr_l^3}{r_s-r_l}\frac{\rho(r_l)}{\bar{M}}\frac{dN}{dM}[m(\thetaE,r_l)]
\label{dtaudtheta}
\end{eqnarray}
where $m(\thetaE,r_l)=(c^2/4G)\thetaE^2 r_s r_l/(r_s-r_l)$.

The optical depth distribution describes the instantaneous properties
of lensing events at any given time, but we are interested in the
distribution of all events.  This is described by the lensing rate
$\Gamma=n\sigma v$.  If the transverse velocity distribution of source
stars is $f_s({\bm v}_s)$, and that of the lenses is $f_l({\bm v}_l)$,
we can write the lensing rate as \citep{griest91}
\begin{equation}
\Gamma=\int dM \frac{dN}{dM} \int_0^{r_s} dr_l
2 r_l\thetaE n(r_l) \bar{v}_T(r_l)
\label{gamma}
\end{equation}
where
\begin{equation}
\bar{v}_T(r_l)=\int d^2v_l d^2v_s f_l({\bm v}_l) f_s({\bm v}_s)
|{\bm v}_T|
\label{vtbar}
\end{equation}
\begin{equation}
{\bm v}_T={\bm v}_l-(1-x){\bm v}_o - x{\bm v}_s.
\end{equation}
Here, ${\bm v}_o$ is the observer's velocity transverse to the line of
sight, and $x=r_l/r_s$.  
We assume all sources are bulge stars, with no net rotation
and a velocity dispersion in each transverse direction of
$\sigma_b$.  Lenses are assumed to reside either in the bulge
or in the disk, with the latter population possessing a flat rotation
curve $v_c$, along with velocity dispersion in each
transverse direction of $\sigma_d$.
For bulge lenses, this gives
\begin{equation}
\bar{v}_T(r_l)=\frac{1}{(1+x^2)\sigma_b^2}
\int v^2 dv \exp\left(-\frac{v^2+(1-x)^2v_c^2}{2(1+x^2)\sigma_b^2}\right)
\frac{1}{2\pi}\int_0^{2\pi}\exp\left(-\frac{(1-x)v_cv}{(1+x^2)\sigma_b^2}
\cos\theta\right)d\theta.
\end{equation}
As usual, the angular integral gives a Bessel function, but note that
it is a modified Bessel function due to the lack of an $i$ in the
exponent.  Thus we have
\begin{eqnarray}
\label{vtran}
\bar{v}_T(r_l)&=&
\frac{\exp\left[-\frac{(1-x)^2v_c^2}{2(1+x^2)\sigma_b^2}
\right]}{(1+x^2)\sigma_b^2}
\int v^2 dv \exp\left(-\frac{v^2}{2(1+x^2)\sigma_b^2}\right)
I_0\left[\frac{(1-x)v_cv}{(1+x^2)\sigma_b^2}\right]\\
&=&\sqrt{\frac{\pi}{2}}\frac{e^{-\frac{(1-x)^2v_c^2}{4(1+x^2)\sigma_b^2}}}
{\left(4(1+x^2)\sigma_b^2 \right)^{1/2}}
\left[\left( (1-x)^2v_c^2+2(1+x^2)\sigma_b^2\right)
I_0\left(\frac{(1-x)^2v_c^2}{4(1+x^2)\sigma_b^2}\right)\right.\nonumber\\
&&\left.\hspace{2in}
+(1-x)^2v_c^2 I_1\left(\frac{(1-x)^2v_c^2}{4(1+x^2)\sigma_b^2}\right)\right]
\nonumber
\end{eqnarray}
For disk lenses, the expression is similar, with $(1-x)v_c\to xv_c$ and
$(1+x^2)\sigma_b^2\to\sigma_d^2+x^2\sigma_b^2$. 

From Eqn.~(\ref{gamma}), the distribution of rate with respect to
angular Einstein radius is
\begin{eqnarray}
\frac{d\Gamma}{d\thetaE}&=&\int dM \frac{dN}{dM} \int_0^{r_s} dr_l
2 r_l\thetaE n(r_l) \bar{v}_T(r_l)
\delta\left[\thetaE-\left(\frac{4GM}{c^2}\frac{r_s-r_l}{r_sr_l}\right)^{1/2}
\right]\nonumber\\
&=&4\int_0^{r_s}dr_l r_l n(r_l) \bar{v}_T(r_l) m(\thetaE,r_l)
\frac{dN}{dM}\left[ m(\thetaE,r_l) \right].
\label{dGdthE}
\end{eqnarray}

Lastly, we also want the joint distribution of rate with respect to
$\thetaE$ and $\tE$.  This is 
\begin{equation}
\frac{d\Gamma}{d\thetaE d\tE}=
4\int_0^{r_s}dr_l r_l n(r_l) m(\thetaE,r_l)
\frac{dN}{dM}\left[ m(\thetaE,r_l) \right]
\int d^2v_l d^2v_s f_l({\bm v}_l) f_s({\bm v}_s)
v_T\delta\left(\tE-\frac{r_l\thetaE}{\tE}\right).
\end{equation}
For bulge lenses, this is
\begin{eqnarray}
\frac{d\Gamma}{d\thetaE d\tE}&=&
4\int_0^{r_s}dr_l n(r_l) m(\thetaE,r_l)
\frac{dN}{dM}\left[ m(\thetaE,r_l) \right]
\frac{r_l^4\thetaE^3}{\tE^4}\nonumber\\&&\qquad\times
\frac{\exp\left[-\frac{(r_l\thetaE/\tE)^2+(1-x)^2v_c^2}{2(1+x^2)\sigma_b^2}
\right]}{(1+x^2)\sigma_b^2}
I_0\left[\frac{(1-x)v_cr_l\thetaE/\tE}{(1+x^2)\sigma_b^2}\right].
\label{dGdthEdtE}
\end{eqnarray}
For disk lenses, the expression is similar, with $(1-x)v_c\to xv_c$ and
$(1+x^2)\sigma_b^2\to\sigma_d^2+x^2\sigma_b^2$.

\end{document}